\documentclass[twocolumn,showpacs,superscriptaddress,prc,aps,10pt,floatfix,nofootinbib]{revtex4-2}
\usepackage{dcolumn,pstricks,amssymb}
\usepackage[intlimits]{amsmath}
\usepackage{graphicx}
\usepackage{bm}
\usepackage{siunitx}
\usepackage[colorlinks=true,linkcolor=black, citecolor=blue]{hyperref}
\newcommand{\tsups}[1]{\textsuperscript{#1}}  
\newcommand{\tsubs}[1]{$_{\textnormal{#1}}$}  

\renewcommand{\vec}[1]{\bm#1}
\usepackage{siunitx}

\definecolor{blue4}{rgb}{0.,0.,0.57917}
\definecolor{green4}{rgb}{0.,0.57917,0.}

\begin{document}

\title{\boldmath First observation of cyclotron radiation from MeV-scale $e^\pm$ following nuclear $\beta$ decay}

\author{W.~Byron}
\affiliation{Department of Physics, University of Washington, Seattle, WA 98195}
\affiliation{Center for Nuclear Physics and Astrophysics, University of Washington, Seattle, WA 98195}

\author{H.~Harrington}
\affiliation{Department of Physics, University of Washington, Seattle, WA 98195}
\affiliation{Center for Nuclear Physics and Astrophysics, University of Washington, Seattle, WA 98195}

\author{R.\,J.~Taylor}
\affiliation{Physics Department, North Carolina State University, Raleigh, NC 27695} 
\affiliation{The Triangle Universities Nuclear Laboratory, Durham, NC 27708}

\author{W.~DeGraw}
\affiliation{Department of Physics, University of Washington, Seattle, WA 98195}
\affiliation{Center for Nuclear Physics and Astrophysics, University of Washington, Seattle, WA 98195}

\author{N.~Buzinsky}
\affiliation{Department of Physics, University of Washington, Seattle, WA 98195}
\affiliation{Center for Nuclear Physics and Astrophysics, University of Washington, Seattle, WA 98195}

\author{B.~Dodson}
\affiliation{Department of Physics, University of Washington, Seattle, WA 98195}
\affiliation{Center for Nuclear Physics and Astrophysics, University of Washington, Seattle, WA 98195}

\author{M.~Fertl}
\affiliation{Institute for Physics, Johannes-Gutenberg University Mainz, 55128 Mainz, Germany}

\author{A.~Garc\'{\i}a}
\affiliation{Department of Physics, University of Washington, Seattle, WA 98195}
\affiliation{Center for Nuclear Physics and Astrophysics, University of Washington, Seattle, WA 98195}

\author{G.~Garvey}
\affiliation{Department of Physics, University of Washington, Seattle, WA 98195}
\affiliation{Center for Nuclear Physics and Astrophysics, University of Washington, Seattle, WA 98195}

\author{B.~Graner}
\affiliation{Department of Physics, University of Washington, Seattle, WA 98195}
\affiliation{Center for Nuclear Physics and Astrophysics, University of Washington, Seattle, WA 98195}

\author{M.~Guigue}
\affiliation{Pacific Northwest National Laboratory, Richland, WA 99352}

\author{L.~Hayen}
\affiliation{Physics Department, North Carolina State University, Raleigh, NC 27695}
\affiliation{The Triangle Universities Nuclear Laboratory, Durham, NC 27708}

\author{X.~Huyan}
\affiliation{Pacific Northwest National Laboratory, Richland, WA 99352}

\author{K.S.~Khaw}
\affiliation{Department of Physics, University of Washington, Seattle, WA 98195}
\affiliation{Center for Nuclear Physics and Astrophysics, University of Washington, Seattle, WA 98195}

\author{K.~Knutsen}
\affiliation{Department of Physics, University of Washington, Seattle, WA 98195}
\affiliation{Center for Nuclear Physics and Astrophysics, University of Washington, Seattle, WA 98195}

\author{D.~McClain}
\affiliation{Department of Physics \& Astronomy, Texas A\&M University, College Station, TX 77843}
\affiliation{Cyclotron Institute, Texas A\&M University, College Station, TX 77843}

\author{D.~Melconian}
\affiliation{Department of Physics \& Astronomy, Texas A\&M University, College Station, TX 77843}
\affiliation{Cyclotron Institute, Texas A\&M University, College Station, TX 77843}

\author{P.~M\"uller}
\affiliation{Physics Division, Argonne National Laboratory, 9700 S. Cass Ave., Argonne, IL 60439}

\author{E.~Novitski}
\affiliation{Department of Physics, University of Washington, Seattle, WA 98195}
\affiliation{Center for Nuclear Physics and Astrophysics, University of Washington, Seattle, WA 98195}

\author{N.\,S.~Oblath}
\affiliation{Pacific Northwest National Laboratory, Richland, WA 99352}

\author{R.\,G.\,H.~Robertson}
\affiliation{Department of Physics, University of Washington, Seattle, WA 98195}
\affiliation{Center for Nuclear Physics and Astrophysics, University of Washington, Seattle, WA 98195}

\author{G.~Rybka}
\affiliation{Department of Physics, University of Washington, Seattle, WA 98195}
\affiliation{Center for Nuclear Physics and Astrophysics, University of Washington, Seattle, WA 98195}

\author{G.~Savard}
\affiliation{Physics Division, Argonne National Laboratory, 9700 S. Cass Ave., Argonne, IL 60439}

\author{E.~Smith}
\affiliation{Department of Physics, University of Washington, Seattle, WA 98195}
\affiliation{Center for Nuclear Physics and Astrophysics, University of Washington, Seattle, WA 98195}

\author{D.D.~Stancil}
\affiliation{Department of Electrical and Computer Engineering, North Carolina State University, Raleigh, NC 27695}

\author{M.~Sternberg}
\affiliation{Department of Physics, University of Washington, Seattle, WA 98195}
\affiliation{Center for Nuclear Physics and Astrophysics, University of Washington, Seattle, WA 98195}

\author{D.\,W.~Storm}
\affiliation{Department of Physics, University of Washington, Seattle, WA 98195}
\affiliation{Center for Nuclear Physics and Astrophysics, University of Washington, Seattle, WA 98195}

\author{H.\,E.~Swanson}
\affiliation{Department of Physics, University of Washington, Seattle, WA 98195}
\affiliation{Center for Nuclear Physics and Astrophysics, University of Washington, Seattle, WA 98195}

\author{J.\,R.~Tedeschi}
\affiliation{Pacific Northwest National Laboratory, Richland, WA 99352}

\author{B.\,A.~VanDevender}
\affiliation{Pacific Northwest National Laboratory, Richland, WA 99352}
\affiliation{Department of Physics, University of Washington, Seattle, WA 98195}

\author{F.\,E.~Wietfeldt}
\affiliation{Department of Physics and Engineering Physics, Tulane University, New Orleans, LA 70118}

\author{A.\,R.~Young}
\affiliation{Physics Department, North Carolina State University, Raleigh, NC 27695}
\affiliation{The Triangle Universities Nuclear Laboratory, Durham, NC 27708}

\author{X.~Zhu}
\affiliation{Department of Physics, University of Washington, Seattle, WA 98195}
\affiliation{Center for Nuclear Physics and Astrophysics, University of Washington, Seattle, WA 98195}

\collaboration{He6-CRES collaboration}
\noaffiliation

\date{\today}
\begin{abstract}
We present an apparatus for detection of cyclotron radiation yielding a frequency-based $\beta^{\pm}$ kinetic energy determination in the 5 keV to 2.1 MeV range, characteristic of nuclear $\beta$ decays. The cyclotron frequency of the radiating $\beta$ particles in a magnetic field is used to determine the $\beta$ energy precisely.
Our work establishes the foundation to apply the Cyclotron Radiation Emission Spectroscopy (CRES) technique, developed by the Project 8 collaboration, far beyond the 18-keV tritium endpoint region.
We report initial measurements of $\beta^-$s from $^6{\rm He}$ and $\beta^+$s from $^{19}{\rm Ne}$ decays to demonstrate the broadband response of our detection system and assess potential systematic uncertainties for $\beta$~spectroscopy over the full (MeV) energy range.
To our knowledge, this is the first direct observation of cyclotron radiation from individual highly relativistic $\beta$s in a waveguide.
This work establishes the application of CRES to a variety of nuclei, opening its reach to searches for new physics beyond the TeV scale via precision $\beta$-decay measurements.
\end{abstract}
\maketitle
Precision measurements of low-energy observables probe the existence of beyond the standard model (BSM) interactions at high energy scales through forces mediated by the exchange of virtual particles with large masses. 
$\beta$-decay spectroscopy provides high intrinsic sensitivity to chirality-flipping interactions not included in the weak sector of the standard model of particle physics~\cite{cirigliano:2013,gonzalez:2018}.
Crucially, distortions to the $\beta$ spectrum depend linearly on BSM scalar and tensor couplings via a characteristic Fierz interference term $b_\mathrm{Fierz}$. Measurements of $b_\mathrm{Fierz}$ at the part-per-thousand level probe new physics above the TeV scale
~\cite{Bhattacharya2012, Falkowski2021, Cirigliano2022}. 

The experimental need for precise $\beta$-decay and conversion-electron spectroscopy goes far beyond searches for scalar and tensor currents, providing a complementary experimental handle for resolving outstanding questions in nuclear physics.
For example, deviations of reactor neutrino spectra from theoretical expectations have been interpreted as evidence of new physics~\cite{hayes:2015,seo:2018,hayen:2019},
while many dark matter searches depend on accurate background modelling of beta radioactivity~\cite{aprile:2020,haselschwardt:2020}. In addition, upcoming studies of solar hep neutrinos~\cite{theia:2022,kub:2004} depend on an accurate determination of the $^8{\rm B}$ neutrino spectrum, which is currently deduced from alpha spectra~\cite{longfellow:2023}. High-precision beta spectroscopy would assist in constraining relevant systematic errors (see, e.g. Ref.~\cite{bahcall:1996} on combining information for the $^8{\rm B}$ spectra.)
Due to the simplicity of the dominant operators, $\beta$ spectroscopy can also be used as a powerful tool to explore nuclei far from stability~\cite{lub:2019} or to help understand stellar nucleosynthesis~\cite{cow:2021}.
High-resolution conversion-electron spectroscopy can also be valuable for studying shape coexistence or other nuclear-structure phenomena~\cite{gar:2020}.

Traditional spectroscopy techniques using semiconductor or scintillator detectors~\cite{UCNA:2020,Sau:2020,Nab:2009,Nab:2019,NOMOS:2019,PERC:2019, naviliat:2018, flechard:2020} that rely on $\beta$ energy loss in matter contend with percent-level corrections due to $\beta$ (back-) scattering and bremsstrahlung losses.
Novel technologies such as neutral atom and ion traps have been employed to reduce these effects, some reaching 0.3\% precision in parameters of the angular distribution~\cite{gorelov-PRL94-2005,BPT,WISArD,fenker-PRL120-2020}. 
Alternative approaches such as superconducting spectrometers~\cite{knutson:2011} and new 
ion traps~\cite{brodeur,tamutrap} are being developed and have shown the potential for similar precision
capabilities~\cite{george:2014,voytas:2015}. 

By contrast, Cyclotron Radiation Emission Spectroscopy (CRES) \cite{p8:proposal}, initially developed by the Project 8 collaboration \cite{p8:prl,p8:prc} for the measurement of the absolute neutrino mass scale from $\beta$-decay spectroscopy of tritium ($\beta^-$ endpoint $\approx 18$~keV), is a promising avenue for low-background and high-resolution ($\delta E/ E \sim 10^{-3}$) $\beta$ spectroscopy.

CRES determines the energy ($E=\gamma m c^2$) of a particle with mass $m$ and charge $q$ by measuring the frequency of the cyclotron radiation emitted in an external magnetic field with magnitude $B$, both of which can be determined to high precision:
\begin{equation}
  f_c = \frac{|q|}{2\pi}\;\frac{B \;c^2}{E}. \label{eqn:cyclotron rad}
\end{equation}

Here we present the first detection of MeV-scale 
$\beta^\pm$s produced by the decays of $^{6}$He ($\beta^-$ endpoint $\approx 3508$~keV) and $^{19}$Ne ($\beta^+$ endpoint $\approx 2216$~keV) with CRES. 

 \tsups{6}He and \tsups{19}Ne are experimentally favorable because both decay almost exclusively to the ground state of their progeny, both have considerably simplified nuclear structure corrections due to an underlying isospin symmetry, and both have precise theoretical characterizations of their decay properties~\cite{naviliat:2009,knecht:2012,hayen:2020,mueller:2022,marcucci:2021,king:2022,glickmagid:2022}.
A measurement of the ratio of $\beta$ energy spectra of these isotopes, only viable given narrow line shapes, enhances the sensitivity to tensor couplings since the sign of $b_\mathrm{Fierz}$ is opposite for $\beta^-$ and $\beta^+$ decays~\cite{jack:1957}.
Systematic effects arising from common spectral distortions are eliminated because CRES has the same response for electrons and positrons since the trajectories and resulting radiation are identical. For example, the spectral ratio cancels out energy-dependent variations in the fiducial volume caused by $\beta$ collisions with the waveguide walls. This cancellation in the ratio works due to the high energy resolution of CRES.

Experimental sensitivity to Fierz interference is optimized by observing $\beta$s over a broad range of energies, spanning from zero to several times the electron mass~\cite{gon:2016}. Here we demonstrate a frequency-based energy determination of $\beta^\pm$s in the 5 keV to 2.1 MeV range, with the capability to reach 5 MeV with the current apparatus. By using a broader radio-frequency (RF) band and by scanning the magnetic field we are able to detect $\approx 100$-times higher energies than previously observed with CRES.

Radioactive and cosmogenic backgrounds are strongly suppressed in CRES experiments, since all charged particles originating from outside the detection volume rapidly terminate on the waveguide walls within a single cyclotron orbit, resulting in no observable cyclotron radiation signal.

\begin{figure}[t]\centering
  \includegraphics[width=0.4\textwidth]{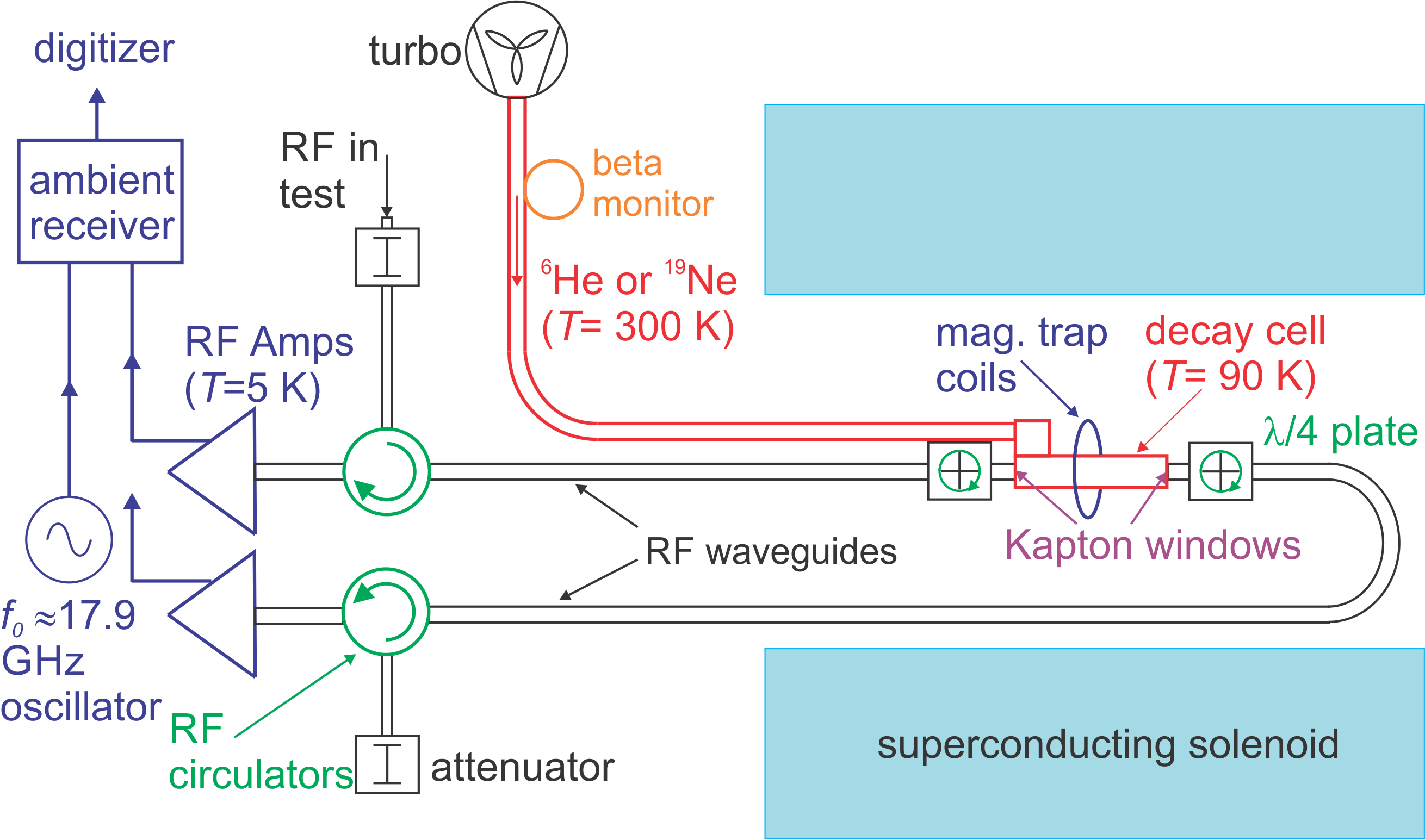}
  \caption{He6-CRES RF setup. The radioactive gases are compressed by a turbo pump into a decay cell with Kapton windows. The decay cell ($T\approx 90~{\rm K}$ to allow for calibrations with $^{83{\rm m}}{\rm Kr}$) is part of an RF waveguide system that transports signals to two low-noise amplifiers ($T \approx 5~{\rm K}$).}
  \label{fig:RF-setup}
\end{figure}

The radioactive isotopes are produced by beams from the FN-tandem accelerator at the University of Washington: via $^7{\rm Li}(d,{^3{\rm He}})^6{\rm He}$ bombarding a molten lithium target with a 17.8-MeV deuteron beam~\cite{knecht:2011} and via $^{19}{\rm F}(p,n)^{19}{\rm Ne}$ 
on an ${\rm SF}_6$ gaseous target with a 12-MeV proton beam, similar to the Berkeley-Princeton source~\cite{dobson:1963,girvin:1972,calaprice:1973,schneider:1983}.
The radioactive atoms are transported to a low-background room through a 15-cm-diameter, 8-m-long stainless-steel pipe, at the end of which a turbo pump is used to compress them into the decay cell, as sketched in Fig.~\ref{fig:RF-setup}. 
Calibrations were performed with $^{\rm 83m}{\rm Kr}$, produced from $^{83}{\rm Rb}$ embedded in zeolites~\cite{ven:2005} and connected to the apparatus.

\begin{figure}[t]\centering
  \includegraphics[width=0.48\textwidth]{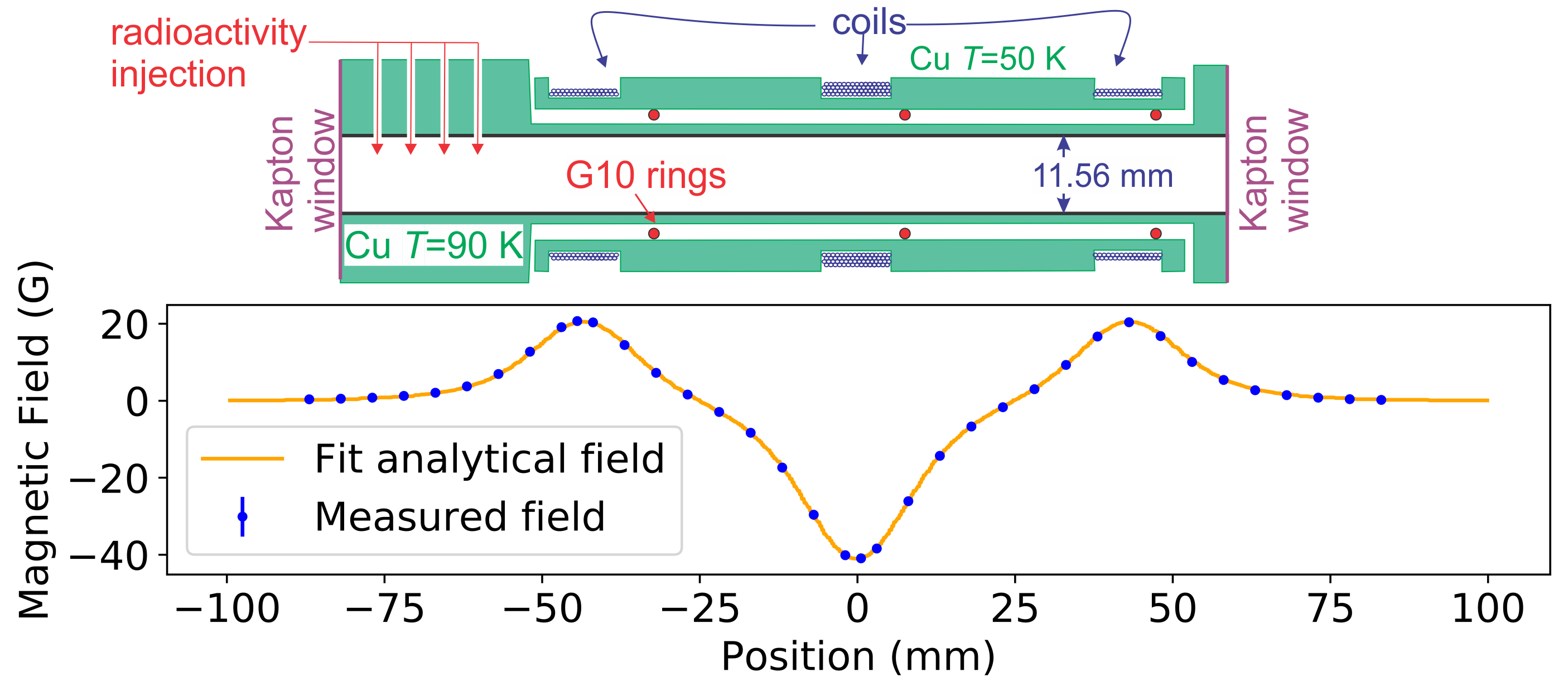}
  \caption{Top: decay cell geometry with trap coils. Bottom: axial magnetic field profile from trap coils used for axial confinement of $\beta$s.}
  \label{fig:Trap Field}
\end{figure}

The decay cell, itself a 11.56-mm diameter circular waveguide, sits inside a superconducting solenoid magnet with a maximum strength of 7 T and uniformity of a few parts per million per cm. This magnet provides the main field along the axis of the waveguide decay cell which causes the $\beta$s to undergo cyclotron trajectories. Three additional coils generate the magnetic trapping field, typically $\mathcal{O}(10^{-3})$ of the main field, used to axially confine the $\beta$s (Fig.~\ref{fig:Trap Field}).

The circularly-polarized cyclotron radiation produced in the decay cell is linearly polarized by $\lambda/4$ phase-retarding polarizers and transported via WR-42 rectangular waveguides to low-noise amplifiers (LNAs). The LNAs~\cite{LNAs:2022} are kept at $\approx 5~{\rm K}$ and the RF noise is $\approx 60~{\rm K}$, dominated by the coupling of the 90-K thermal noise in the waveguide. 
Signals from the LNAs are mixed with a 17.9~GHz reference oscillator in a heterodyne receiver at room temperature, down-converting signal frequencies from 18--19.1~GHz to 0.1--1.2~GHz. This is much broader than the 180-MHz total bandwidth used to measure the tritium spectrum with CRES~\cite{p8:tritium}. A ROACH2 system~\cite{roach2} digitizes the time-series data at 2.4~GHz in $\approx 6.8\, \mu{\rm s}$ segments, converting each segment into the frequency domain by fast Fourier transform (FFT), which is then recorded at $\approx 600~{\rm MB/s}$.

\begin{figure*}[t]
  \hfil\scalebox{0.21}{\includegraphics*[trim={10 0 0 0},clip]{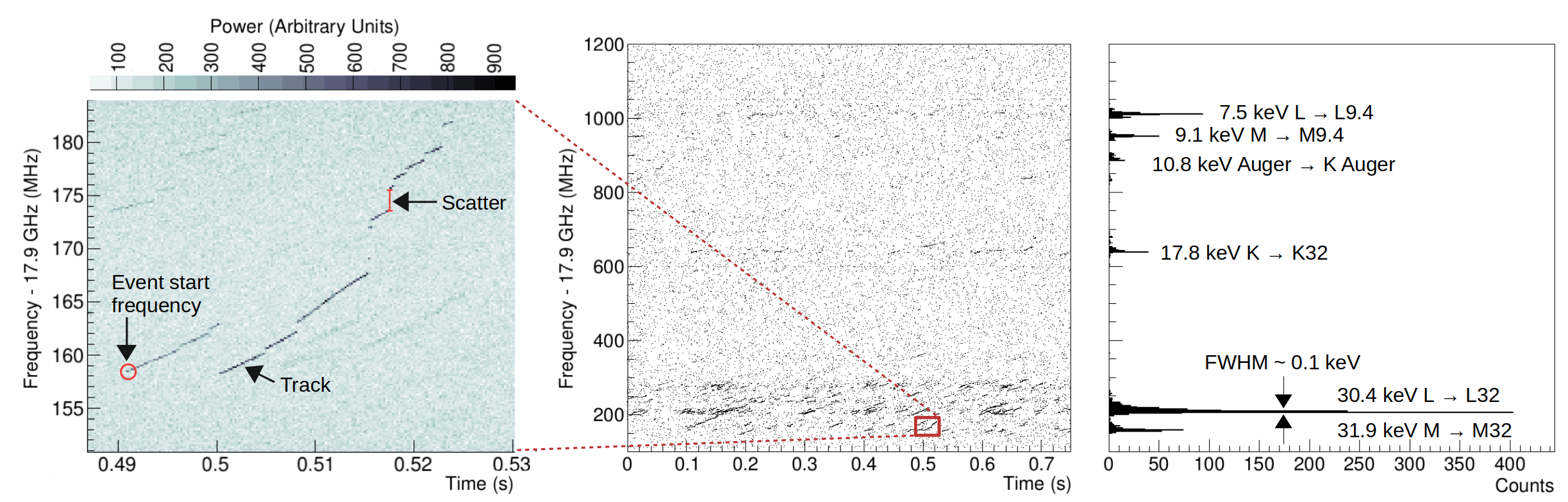}}\hfil
  \caption{$^{83{\rm m}}{\rm Kr}$ data taken with $B=0.68$~T. Center: spectrogram with Fourier bin threshold ${\rm SNR}>6$ (black), demonstrating the full 1.1~GHz bandwidth. Left: zoomed-in region exemplifying an event composed of multiple tracks. Right: reconstructed event start frequency histogram, showing simultaneous observation of the 7--32 keV lines.}
  \label{fig:83Kr-spectrogram}
\end{figure*}

Arranged sequentially, the data produces a \textit{spectrogram} displaying signal power as a function of time and frequency.
Figure~\ref{fig:83Kr-spectrogram} shows CRES data taken with the $^{83{\rm m}}{\rm Kr}$ source.
Consecutive and colinear high-power Fourier bins form \textit{tracks}, which are indicative of a radiating charged particle.
The gradual increase of frequency versus time is due to the $\beta$ losing energy to the cyclotron radiation (Eq.~\ref{eqn:cyclotron rad}).
The sudden jumps are due to scattering off residual atoms.
The pressure in the decay cell, achieved via getter and cryo pumps, was $\approx 10^{-7}$~mbar, and dominated by residual hydrogen and nitrogen.
The set of all tracks associated with a single $\beta$, which may have an arbitrary number of scatters, is referred to as an \textit{event}.

To robustly reconstruct events our data analysis involves (1) applying a signal-to-noise-ratio (SNR) cut, (2) clustering points into tracks, and (3) clustering tracks into events. Katydid~\cite{katydid}, a modular C\texttt{++}-based analysis framework developed by the Project 8 collaboration, has been adapted for the specific event reconstruction needs of He6-CRES. The quality of event reconstruction was verified using both visual inspection and comparison of expected and observed distributions of reconstructed event properties.

Event start frequencies and magnetic field strengths are used to determine $\beta$ energies using Eq.~\ref{eqn:cyclotron rad}, which are histogrammed into an experimental $\beta$ spectrum.
CRES energy resolution scales with event start frequency resolution as $\delta E/E \approx \delta f_c /f_c$ implying that 20 MHz frequency resolution at $f_c\approx$ 20 GHz, readily attainable in He6-CRES across the 5 keV -- 2.2 MeV energy bandwidth, yields part-per-thousand energy resolution.

\begin{figure*}[t]
 \hfil\scalebox{0.37}{\includegraphics*{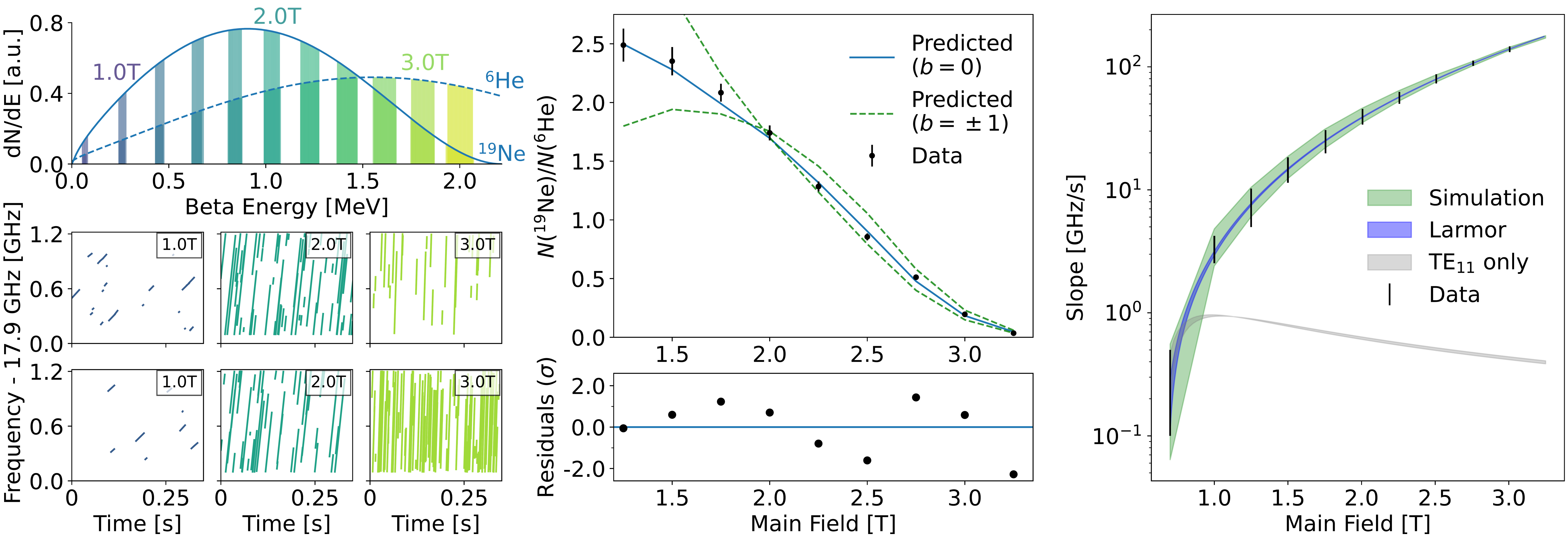}}\hfil
  \caption{Summary of CRES events from $^{19}{\rm Ne}$ and $^{6}{\rm He}$.
    Top left: Standard model $^{19}{\rm Ne}$ and $^{6}{\rm He}$ $\beta$ spectra showing the energy regions sampled by the 11 field settings from 0.75--3.25 T, given the 18--19.1-GHz RF bandwidth.
  Bottom Left: Grid of identified tracks from $^{19}{\rm Ne}$ (top) and $^{6}{\rm He}$ (bottom) at 1, 2, 3 T.
    Each panel is an overlay of 0.35 seconds of data from 5 separate acquisitions.
   Track colors correspond to the magnetic field settings from the top left plot.
Center: Spectral ratio plot of $^{19}{\rm Ne}/{}^{6}{\rm He}$ observed with CRES, including residuals with respect to Monte Carlo (lower inset). Different scattering environments for the two isotopes prevent the cancellation of efficiencies at low fields (not shown). Right: Observed track slopes ($df_c/dt$) versus magnetic field for $^{19}{\rm Ne}$ and $^{6}{\rm He}$, showing agreement with numerical waveguide simulations (green) and the Larmor formula (blue), and comparing to coupling to TE$_{1,1}$ mode only (gray).}
  \label{fig: spectral ratio}
\end{figure*}

Figure~\ref{fig: spectral ratio} (top left) illustrates how the $\beta$ spectrum of $^{19}{\rm Ne}$ and $^{6}{\rm He}$ are covered by scanning the magnetic field, given a fixed frequency bandwidth.
Measurements taken at each field are normalized by a $\beta$ monitor of the total activity within the decay cell.
Shown in Fig.~\ref{fig:RF-setup}, a 3.8-cm-diameter port with a 1/8-mm thick Cu foil allows $\beta$s to be observed by a 5-mm-thick scintillator coupled to 4 SiPM readouts. The density of radioactive atoms in the vacuum side of this port is in dynamical equilibrium with that in the decay cell.

Event features vary significantly with $\beta$ energy. Most prominently, high energy $\beta$s result in long duration and highly sloped tracks (Fig.~\ref{fig: spectral ratio}, bottom left).

Track durations are limited by the time interval between scatters with the residual gas. Due to the decrease in the scattering cross section for $\beta$s on residual gas atoms with increasing $\beta$ energy the mean track length grows by over a factor of 20 over the observed energy range from 5 keV -- 2.1 MeV.
Longer tracks make identifying the beginning of an event easier, since scattering can lead to the misidentification of the initial frequency. 
The low scattering probability also leads to accumulation of $\beta$s in the trap. These result in a high density of long tracks crossing the spectrogram, interfering with the correct reconstruction of events in our bandwidth. To remedy this, the trapping coils are toggled on and off with a period of 55 ms, effectively emptying the trap.
As cycling the trapping fields adversely affects the duty cycle, we plan to reduce the trap-off time to below 1 ms with a redesigned decay cell with thinner walls to mitigate eddy currents, which is the limiting factor. 

An additional difference observed between the tracks for high-energy $\beta$s compared to those from $^{83{\rm m}}{\rm Kr}$ is that the slopes are up to three orders of magnitude larger (Fig.~\ref{fig: spectral ratio}, right).
In a waveguide, the power radiated into a mode $\lambda$ at frequency $f$ is given by~\cite{jackson:1991,pozar:2005,p8:prc}:
\begin{eqnarray}
  P_{\lambda}(f) \propto \left\lvert \int q \vec{v}(t)\cdot\vec{E}_{\lambda}(\vec{r}(t)) e^{-2 \pi i f t} \, dt \right\rvert^2  \label{eq:coupling}
\end{eqnarray}
where $\vec{r}(t), \vec{v}(t)$ are the position and velocity vectors of the moving $\beta$, and $\vec{E}_{\lambda}(\vec{r})$ represents the mode electric field. 
The bandwidth (18--19.1~GHz) was selected to measure only in the TE\tsubs{1,1} mode, with cutoff at 15.2~GHz. The next mode, TM\tsubs{0,1}, has a cutoff at 19.9~GHz. Thus the track power observed is exclusively due to the power propagating in the TE\tsubs{1,1} mode. The track slope, $df_c/dt$, is a measure of the $\beta$'s total radiated power.
In addition to the TE\tsubs{1,1} mode, the $\beta$ velocity vector couples to higher-order waveguide modes at higher harmonic frequencies ($f = n f_c$), which can contribute significantly to the total radiated power and thus to the observed slope (Fig.~\ref{fig: spectral ratio}, right). However, since this power propagates into unobserved modes, we do not detect this power directly.  

Notably, experimentally observed slopes scale in accordance with the expectation for radiation in free space, given by the relativistic Larmor formula~\cite{jackson:1991}.
While in general, the radiated power emitted in a waveguide can vary significantly from the free-space expectation, this result is predicted analytically for orbits centered around the waveguide axis~\cite{dok:2001} and is confirmed numerically using Equation~\ref{eq:coupling}, calculated up to 800 harmonics, as shown in Fig.~\ref{fig: spectral ratio} (right, green band).

These two track features (long durations and steep slopes) characterize events from high-energy $\beta$s, a novel energy regime for CRES.
The event reconstruction described above was designed to identify start frequencies across the $\beta$ spectrum, and thus for a broad range of track slopes, durations, and signal-to-noise ratios.

Figure~\ref{fig: spectral ratio} (center) presents a preliminary ratio of \tsups{19}Ne and \tsups{6}He experimental event rates compared to
Monte Carlo predictions given $b_\textrm{Fierz}=0, \pm 1$. Monte Carlo simulations account for events that originate below the visible frequency bandwidth and rise into our observation window using the known expected track slope and magnetic trapping period.
No event start frequency cut was applied to data to mitigate systematic uncertainties resulting from imperfect vetoing of events originating from below the frequency bandwidth.
Systematic uncertainties were estimated by calculating the range of ratios obtained from different Fourier bin SNR thresholds (8, 9, 10, 11) and added in quadrature with the statistical uncertainties.

Residual gases in the decay cell were different in the \tsups{19}Ne and \tsups{6}He measurements leading to statistically distinct track-duration distributions below 1.0 T. Scattering systematics are no longer cancelled in the spectral ratio at these fields.

The application of CRES to precision broadband energy spectroscopy is a promising avenue for searching for new physics.
Excellent energy resolution and symmetric detector responses between $\beta^\pm$s make CRES well suited to high precision spectral ratio measurements. The technique can be extended to other isotopes, including ion beams of short-lived, exotic nuclei produced at specialized facilities, such as FRIB \cite{FRIB:2023}.

The spectral ratio in Fig.~\ref{fig: spectral ratio} contains $10^4$ counts per isotope obtained from 90 minutes of CRES data each.
Discovery-level sensitivity to $b_\textrm{Fierz}$ with CRES requires larger event rates via higher source intensity or detection efficiency. An immediate order of magnitude increase in the event rate is expected from improving radioactive isotope transport and trapping duty cycles. Presently the CRES detection efficiency is $\lesssim 0.1\%$ due to trapping inefficiencies and low SNR events, in part due to the large variations in track slopes described above. RF hardware alterations targeting these SNR fluctuations and analysis improvements should allow for another factor of 5 increase in CRES rate. With these upgrades, statistical sensitivity of $b_{\textrm{Fierz}}\sim 10^{-3}$ could be obtained with $\approx 10$ days of CRES data.

In summary, we present the first observation of CRES events for $\beta$s from $^6{\rm He}$ and $^{19}{\rm Ne}$ decays. Using a broad RF bandwidth and scanning the magnetic field, our apparatus can measure $\beta$s with kinetic energies in the $5~{\rm keV}$ to $5~{\rm MeV}$ range. The radiated power emitted by $\beta$s in the waveguide was shown to present wide variations, according to the field, but this and other effects can be mitigated by spectral ratio measurements. This Letter therefore establishes the practical application of the CRES technique to a variety of nuclei covering the range of typical nuclear $\beta$-decay energies. This capability enables high precision determinations of $\beta$ spectra that can be used to search for signatures of BSM physics above the TeV scale in addition to serving as a useful complementary tool for other applications.

\acknowledgements
We thank Andre Young, Christine Claessens, Ali Ashtari Esfahani, Gerald Miller for useful guidance and discussions, David Peterson, Tim Van Wechel, and Ryan Roehnelt for help with the apparatus and the CASPER collaboration for help with the ROACH digitizers and signal processing. 
This work is supported by the US Department of Energy (DOE), Office of Nuclear Physics, under Contracts No.~DE-AC02-06CH11357, DE-FG02-97ER41020, DE-FG02-ER41042, DE-AC05-76RL01830, DE-FG02-93ER40773, by the National Nuclear Security Administration under Award No.~DE-NA0003841, by the National Science Foundation, grants No. NSF-1914133 and  PHY-2012395, by the Gordon and Betty Moore Foundation and by the Cluster of Excellence “Precision Physics, Fundamental Interactions, and Structure of Matter” (PRISMA+ EXC 2118/1) funded by the German Research Foundation (DFG) within the German Excellence Strategy (Project ID 39083149). The $^{83}{\rm Rb}-^{83}{\rm Kr}$ source used in this research was supplied by the DOE by the Isotope Program in the Office of Nuclear Physics.

\bibliography{He6-CRES}
\end{document}